\documentclass[preprint,3p,times,twocolumn]{elsarticle}

\usepackage{amssymb}

\usepackage{amsmath,amsfonts}
\usepackage{algorithmic}
\usepackage{algorithm}
\usepackage{array}
\usepackage[caption=false,font=normalsize,labelfont=sf,textfont=sf]{subfig}
\usepackage{textcomp}
\usepackage{stfloats}
\usepackage{url}
\usepackage{verbatim}
\usepackage{graphicx}

\usepackage[nohyperlinks, nolist]{acronym}
\usepackage[hidelinks]{hyperref}
\usepackage{xspace}
\usepackage{multirow}

\usepackage{orcidlink}

\journal{Sustainable Energy, Grids and Networks}

\begin{document}

\newcommand{\eg}{e.g.\xspace}
\newcommand{\ie}{i.e.\xspace}
\newcommand{\etal}{et al.\xspace}

\newcommand{\pandapower}{\textit{pandapower}\xspace}
\newcommand{\simbench}{\textit{SimBench}\xspace}

\newcommand{\pandapowerfootnote}{\footnote{\url{https://pandapower.readthedocs.io/en/latest/}, last access: 2023-10-30}\xspace}
\newcommand{\simbenchfootnote}{\footnote{\url{https://simbench.de/en/}, last access: 2023-10-30}\xspace}

\newcommand{\power}{P}
\newcommand{\price}{p}
\newcommand{\agentidx}{a}
\newcommand{\nagents}{|\agents|}
\newcommand{\agents}{A}
\newcommand{\genidx}{\agentidx}
\newcommand{\ngens}{\nagents}
\newcommand{\marginal}{\price^\mathrm{marginal}}
\newcommand{\slack}{\mathrm{slack}}
\newcommand{\total}{\mathrm{total}}
\newcommand{\pricerange}{[0, \price_\mathrm{max}]}
\newcommand{\penaltyOPF}{\Psi}
\newcommand{\voltage}{u}
\newcommand{\bus}{b}
\newcommand{\buses}{B}
\newcommand{\branch}{l}
\newcommand{\branches}{L}
\newcommand{\loadidx}{l}
\newcommand{\loads}{L}
\newcommand{\trafo}{t}
\newcommand{\trafos}{T}
\newcommand{\regret}{\psi}  
\newcommand{\totalregret}{\Psi} 
\newcommand{\bestreward}{\reward^*}
\newcommand{\reward}{r}
\newcommand{\obs}{obs}
\newcommand{\act}{act}

\newcommand{\ntest}{50}
\newcommand{\nseed}{10} 
\begin{acronym}[JSONP]\itemsep0pt
	\acro{DER}{distributed energy resource}
	\acro{DSO}{distribution system operator}
	\acro{EHV}{extra high voltage}
	\acro{HV}{high voltage}
	\acro{LV}{low voltage}
	\acro{MV}{medium voltage}
	\acro{OPF}{optimal power flow}
	\acro{PV}{photovoltaic}
	\acro{RES}{renewable energy resources}
	\acro{TSO}{transmission system operator}
	\acro{WT}{wind turbine}
	\acro{RL}{reinforcement learning}
	\acro{DRL}{deep RL}
	\acro{MARL}{multi-agent \acl{RL}}
	\acro{NE}{Nash-equilibrium}
	\acro{MAPE}{mean absolute percentage error}
	
	\acro{NN}{artificial neural network}
	\acro{ML}{machine learning}
	\acro{MADDPG}{Multi-Agent Deep Deterministic Policy Gradient}
	\acro{MMADDPG}[M-MADDPG]{model-extended MADDPG}
    \acro{MATD3}{Multi-Agent Twin-Delayed \ac{DDPG}}
	\acro{DDPG}{Deep Deterministic Policy Gradient}
	\acro{DDQN}{Double Deep Q Network}
	\acro{TD3}{Twin-Delayed \ac{DDPG}}
	\acro{MDP}{Markov decision process}
	\acro{WoLF-PHC}{win or learn fast policy hill climbing}

\end{acronym} 

\begin{frontmatter}



\title{Approximating Energy Market Clearing and Bidding With Model-Based Reinforcement Learning}


\author[label1]{Thomas Wolgast\corref{cor1}\orcidlink{0000-0002-9042-9964}}\ead{thomas.wolgast@uni-oldenburg.de}
\author[label1]{Astrid Nieße\orcidlink{0000-0003-1881-9172}}

\affiliation[label1]
            {organization={Digitalized Energy Systems, University of Oldenburg},
            addressline={Ammerländer Heerstraße 114-118}, 
            city={Oldenburg},
            postcode={26129},
            country={Germany}}

\cortext[cor1]{Corresponding Author}

\begin{abstract}
Energy market rules should incentivize market participants to behave in a market and grid conform way. However, they can also provide incentives for undesired and unexpected strategies if the market design is flawed. \Ac{MARL} is a promising new approach to predicting the expected profit-maximizing behavior of energy market participants in simulation. 
However, \acl{RL} requires many interactions with the system to converge, and the power system environment often consists of extensive computations, \eg, \ac{OPF} calculation for market clearing.
To tackle this complexity, we provide a model of the energy market to a basic \ac{MARL} algorithm in the form of a learned \ac{OPF} approximation and explicit market rules. The learned \ac{OPF} surrogate model makes an explicit solving of the \ac{OPF} completely unnecessary. 
Our experiments demonstrate that the model additionally reduces training time by about one order of magnitude but at the cost of a slightly worse performance. 
Potential applications of our method are market design, more realistic modeling of market participants, and analysis of manipulative behavior.\footnote{URL to all the source code will be added upon publication}


\end{abstract}




\end{frontmatter}


\section{Introduction}
%
To improve competition, economic efficiency, and transparency, the energy system more and more transforms into a market-based system. This trend can be seen in the emerging or changing market designs of, \eg, local energy markets and ancillary service markets, both in scientific literature and in practice.
To design efficient and robust markets,
it is important to predict and understand how rational profit-maximizing agents will behave under a given set of market rules. 
For these kinds of analyses, often simplifying assumptions like the absence of market power are used, \ie, the ability of a participant to drive the price over a competitive level \cite{pra13}. 
While valid in markets with lots of participants, this assumption is questionable in local energy or ancillary service markets with few participants who can be located in strategically advantageous positions. 
It has been shown that -- even in the absence of market power -- grid-harming behavior of the market participants is not only possible but sometimes profitable and therefore considered rational \cite{wol21}: 
One example is the well-understood inc-dec gaming, where market participants adjust their bidding on the wholesale energy market to create or amplify grid congestions. Then, they generate profit by providing ancillary service countermeasures \cite{lio19}. Wolgast \etal \cite{wol21} found a similar manipulation strategy in reactive power markets by using \ac{RL}. The learning agent autonomously learned to attack the system with controllable loads to artificially increase reactive power demand to then profit from its delivery. The only objective of the agent was to maximize profit; the grid-harming behavior emerged as a side-effect. 
\par 
If markets with such unwanted incentives are brought into the field, the potential consequences for grid stability, security of supply, and overall efficiency could be dramatic. It is essential to develop methods to foresee and understand the expected behavior of the market participants during market design.
%
\par 
One recent approach to determine realistic market behavior is by empirically learning individual strategies of the players with \acf{MARL}, \eg by Du \etal \cite{du21} and Rashedi \etal \cite{ras16}.
In \ac{RL} and its multi-agent variant, an environment is required that defines the optimization problem, in this case, profit maximization on the energy market. 
However, often, an \acf{OPF} or other optimization problems are required to solve for the market clearing in the environment.  In \ac{RL}, thousands or millions of interactions with the environment are required, resulting in the same amount of \acp{OPF} to solve and requiring extensive computation. That limits applicability to complex real-world scenarios.

Du \etal and Rashedi \etal use model-free approaches in their publications, which is the most common way in \ac{RL} research \cite{sch20}.
Model-free \ac{RL} algorithms are applied to an environment that holds all the information about the problem to solve. The agent learns the problem from scratch, making the approach generally applicable to various problems.
In contrast, we explicitly integrate domain knowledge into the training to overcome the computational challenge of solving the \ac{OPF} in each step.
Such model-based \ac{RL} algorithms are not applicable to a wide class of problems anymore but are tailored to a specific problem. However, a speed-up of training or improved performance can be expected in domain-specific tasks \cite{yan20c}.
\par 
As our main contribution, we demonstrate how a learned surrogate model can replace the \ac{OPF} for market clearing, which speeds up training by about one order of magnitude and renders a separate solving of the \ac{OPF} completely unnecessary. For that, we discuss three general concepts of how to use domain knowledge to improve training for \ac{MARL} bidding in energy market environments and similar problems.  
\par 
Note that determining the optimal bidding strategy of each player, considering the optimal bidding strategies of all competitors, is to search for the \ac{NE}. However, since guaranteeing convergence of \ac{MARL} to the \ac{NE} is an unsolved problem, we focus on \textit{realistic} market behavior instead. However, using the \ac{NE} as a reference is a useful criterion to evaluate the success of the applied methodology, which we will discuss in section~\ref{sec:regret_test} again. 
%
%
\par 
The remainder is structured as follows. In section~\ref{sec:related}, we present the related work of learning multi-agent bidding in energy markets and approximating the \ac{OPF} with \ac{RL}. In section~\ref{sec:method}, we first present the basic \ac{MADDPG} algorithm and then propose three general ideas to convert it into a model-based \ac{MARL} algorithm. In sections~\ref{sec:scenario} and \ref{sec:implement}, we present an exemplary energy market bidding scenario as \ac{RL} environment and discuss how the model-based ideas can be applied to that specific scenario. In section~\ref{sec:experimentation}, we discuss the experimentation setting, followed by the results in section~\ref{sec:results}. We discuss our results in section~\ref{sec:discussion}, followed by a short conclusion.

\section{Related Work}\label{sec:related}

In the following, we first present the current state of the art of using \ac{MARL} to determine expected bidding behavior and strategies in energy market environments. Further, we present recent related work for approximating the \ac{OPF} for market clearing with \ac{ML} methods.
\subsection{Multi-Agent Bidding in Energy Markets}
Ye \etal \cite{ye19} apply Deep Policy Gradient to train 10 agents to bid in a network-constrained economic dispatch.
They approximate a known \ac{NE}, and the training was even faster than the non-\ac{RL} baseline algorithm. However, their training data only spanned one day.
\par 
Liang \etal \cite{yan20d} use \ac{DDPG} to train up to 24 agents and focus primarily on tacit collusions, \ie implicit price agreements without communication. Their market clearing algorithm maximizes social benefit under line load constraints, similar to an economic dispatch. As Ye \etal, they reproduce a known \ac{NE} with their \ac{RL} method. However, they consider simplified scenarios w.r.t. size and parametrization (max. six agents, constant load and generation). 
\par 
Rashedi \etal \cite{ras16} argue that single-agent \ac{RL} is insufficient for the multi-agent bidding problem and that \ac{MARL} methods are required instead. They apply multi-agent q-learning to learn the optimal bidding behavior of six agents and demonstrate that it outperforms single-agent learning.
The market clearing is done with a security-constrained DC-\ac{OPF} that minimizes economic costs. 
\par 
Gao \etal \cite{gao21} apply the \ac{WoLF-PHC} \ac{MARL} algorithm to the bidding behavior of electric vehicle and wind turbine oligopolies in a stochastic game. Up to 12 agents learn to optimize their bidding behavior; other generators are assumed to be non-strategic. They use data from a single day for training, and the market is cleared with a constrained DC \ac{OPF}.
In follow-up work, Zhu \etal \cite{zhu21} use the same algorithm to identify load demand, congestions, and price caps as key factors that affect the converged bidding strategies.
\par 
Du \etal \cite{du21} formulate the multi-agent bidding problem as a Markov game and apply \ac{MADDPG} to approximate a \ac{NE}.
However, they consider only three agents and assume that all other non-learning agents bid truthfully, which does not hold with strategic bidding.
\par 
Harder \etal \cite{har23} train 25 \ac{MATD3} agents to optimize the bidding behavior of energy storage units in the electricity market. They focus on the scalability of the \ac{MARL} approach and on markets that do not require any optimization for market clearing. 
\par 
Overall, the current state-of-the-art literature has three main drawbacks: 
1)~Most of the scenarios are quite simple, with more or less constant parameters. This way, the \ac{RL} agents are potentially able to memorize the data and can only find their optimal actions for the specific cases used for training, without learning a strategy that generalizes to unseen data. 
2)~The approaches use the binary classification of either determining a \ac{NE} or not (except \cite{har23}).
However, \acp{NN} are only capable of approximation, and especially \ac{DRL} is often quite noisy. Therefore, an exact hit of the equilibrium point is unlikely and also not needed for real-world applications, especially if we attempt to consider more complex scenarios in the future.
3)~In most cases, some optimization problem needs to be solved for market clearing. Since thousands and millions of training steps are typical in \ac{DRL}, this again will cause problems in future complex scenarios because computation times will explode.
\subsection{Learning the Optimal Power Flow}\label{sec:rl-opf} 
The main contribution of our work is to use a learned surrogate model of the \ac{OPF} to train the market participant agents. In recent years, a large body of literature emerged to use \ac{ML} to approximate the \ac{OPF}.
For the sake of brevity, the following overview focuses on selected publications on \ac{ML} for economic dispatch and market clearing, which are especially relevant for this work.
\par 
Duan \etal \cite{dua19b} use a \ac{DDQN} to solve the optimal active power dispatch.
The objective is to minimize active power costs under consideration of voltage constraints. However, they use fixed active power prices, which is unsuitable in a market setting with changing prices. 
\par 
Chen \etal \cite{che22} use supervised learning to approximate a large-scale security-constrained economic dispatch. 
The cost coefficients are part of the \ac{NN} input and get sampled from a fixed data set. They achieved a speed-up of four orders of magnitude with only minimal error.
\par 
Zhen \etal \cite{hon21} model the economic dispatch as 1-step \ac{MDP}, \ie, the solution is not generated iteratively, but in a one-shot fashion. They use the \ac{TD3} algorithm to learn to minimize generation costs under multiple constraints.
\par
Zheng \etal \cite{zhe21} again train a \ac{NN} to minimize generation costs under constraints. They use no standard \ac{DRL} algorithm but derive the gradient directly from several perturbed interactions with the environment. 
They compare their approach to supervised training of the \ac{NN} and achieve significant performance improvement. 
\par 
The previous overview reflects a current research trend to approximate the \ac{OPF} and the economic dispatch with \ac{DRL}. These publications demonstrate that approximation of the \ac{OPF} with \ac{RL} and \acp{NN} is well possible and results in significant speed-up. That happens by transforming the complex non-linear optimization into fast matrix multiplications.
The speed-up is the main motivation. We argue that using the \ac{RL}-\ac{OPF} approximation as a surrogate model for training agents in that environment is a great application for this technique.
%

\section{Multi-Agent Bidding in Energy Markets}\label{sec:method}

Our \ac{MMADDPG} algorithm builds upon the \ac{MADDPG} algorithm, the multi-agent variant of the broadly used \ac{DDPG}. We first briefly introduce \ac{DDPG} and \ac{MADDPG}. Then, we discuss our model-based extensions to tailor \ac{MADDPG} for the energy market bidding problem.
\subsection{Deep Deterministic Policy Gradient (DDPG)}
\ac{DDPG} \cite{lil15} is an actor-critic \ac{DRL} algorithm. The general idea is to train a \textit{critic} \ac{NN} that predicts an action's expected long-term reward, \ie, the Q-value. The \textit{actor} \ac{NN} maps observations to actions, which are fed as input into the critic. Then, by backpropagating through both critic and actor \acp{NN}, we can compute the gradients of the actor weights that maximize the output of the critic, \ie, the long-term reward $Q$. Applying these gradients to the actor improves the agent's performance. 
To improve sample efficiency, all collected samples are stored in a replay buffer, from which the training data gets sampled. 
Since the actor \ac{NN} generates deterministic actions, noise is added to the actions to improve exploration.
For the sake of brevity, we omitted algorithmic details. For a more thorough explanation of \ac{DDPG}, refer to \cite{lil15}.
%
%
%
\subsection{Multi-Agent DDPG (MADDPG)}
Single-agent \ac{RL} algorithms like \ac{DDPG} are not applicable to multi-agent problems, due to the combinatorial complexity, the non-stationary task, and the multi-dimensional learning objectives \cite{yan20c}.
\par 
\ac{MADDPG} \cite{low17} expands \ac{DDPG} to solve cooperative, competitive, or mixed multi-agent problems. 
The general idea remains the same, but now each agent is represented by an actor-critic combination. The critics are trained again to predict the Q-value of their respective agents since each agent has its own goal. As mentioned before, multiple learning agents lead to non-stationary problems. For example, the agent cannot predict its Q-value without knowing the actions of all other agents because it remains unclear who the originator of a good/bad reward was. \ac{MADDPG} solves that problem with centralized training and decentralized execution, which assumes that all agents know all actions and observations of the other agents during training, even in competitive scenarios. This way, training the critic is possible again. However, the actors receive only local observations, so decentralized execution after training is still possible. Note that the centralized training does not result in any privacy issues. We aim to compute expected bidding strategies under given market rules, \ie how would a market player realistically act in some situation? The actual market player does not need to share any actual data to achieve that.  
\par
\begin{figure}
\centering
\includegraphics[width=85mm]{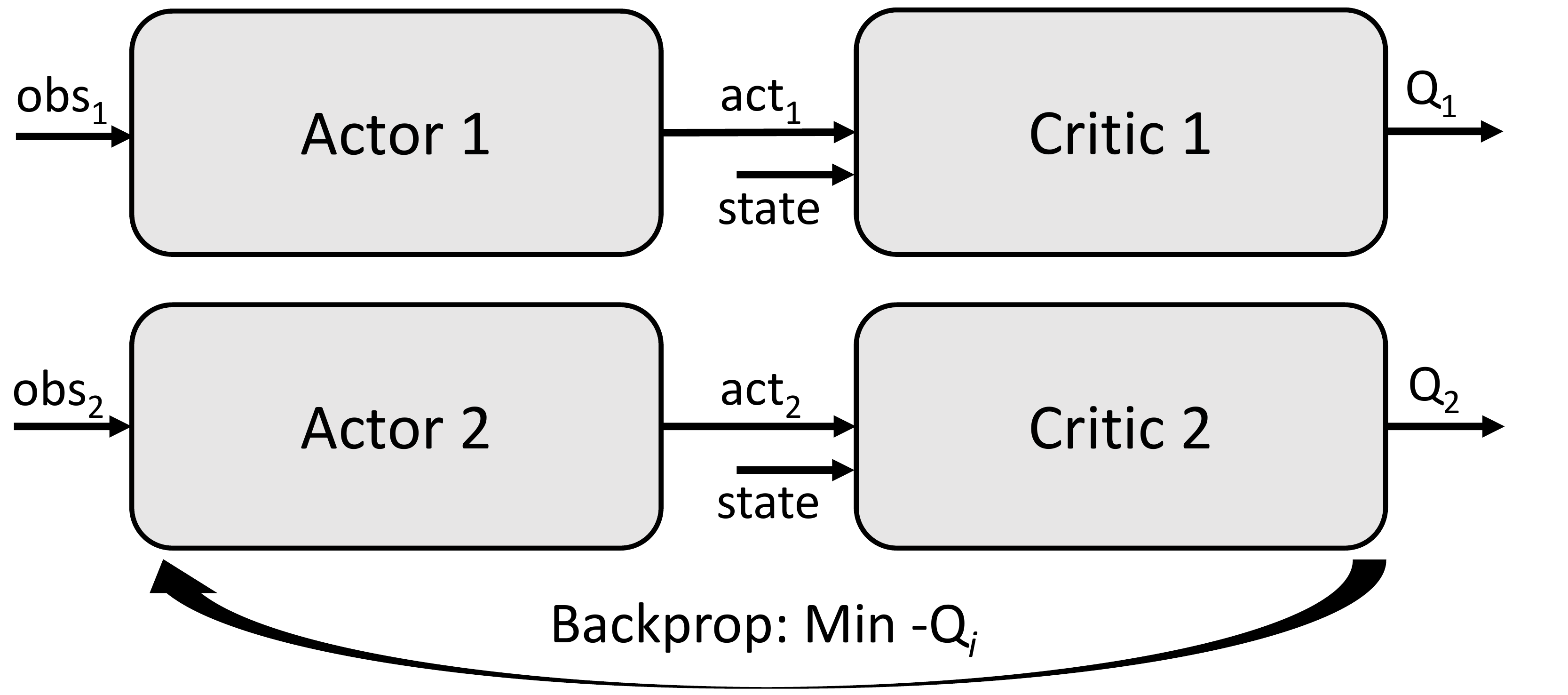}

\caption{Example of \ac{MADDPG} algorithm with two agents.}
\label{fig:maddpg}
\end{figure}

 
Figure~\ref{fig:maddpg} visualizes \ac{MADDPG} by the example of two agents. For a complete description of \ac{MADDPG}, refer to \cite{low17}.

\subsection{Model-Extended MADDPG (M-MADDPG)}\label{sec:mmaddpg}
As mentioned, the \ac{MARL} bidding problem requires solving an \ac{OPF} per environment interaction of the agents. Since more complex \ac{RL} problems require millions of samples, training becomes computationally very heavy. This problem gets aggravated by the number of agents in \ac{MARL} scenarios since the interdependencies result in slower training. 
\par 
To speed up training and potentially improve performance, we introduce three concepts of how \ac{RL} algorithms can be tailored for problems in the domain of energy market bidding. We discuss and demonstrate these concepts by the example of \ac{MADDPG}. However, they are not limited to this specific algorithm and use case, which we will discuss in detail for every concept respectively. In this section, we present only the general concepts. The specific implementation for our scenario is discussed later in section \ref{sec:implement}.  


\begin{figure}
\centering
\includegraphics[width=85mm]{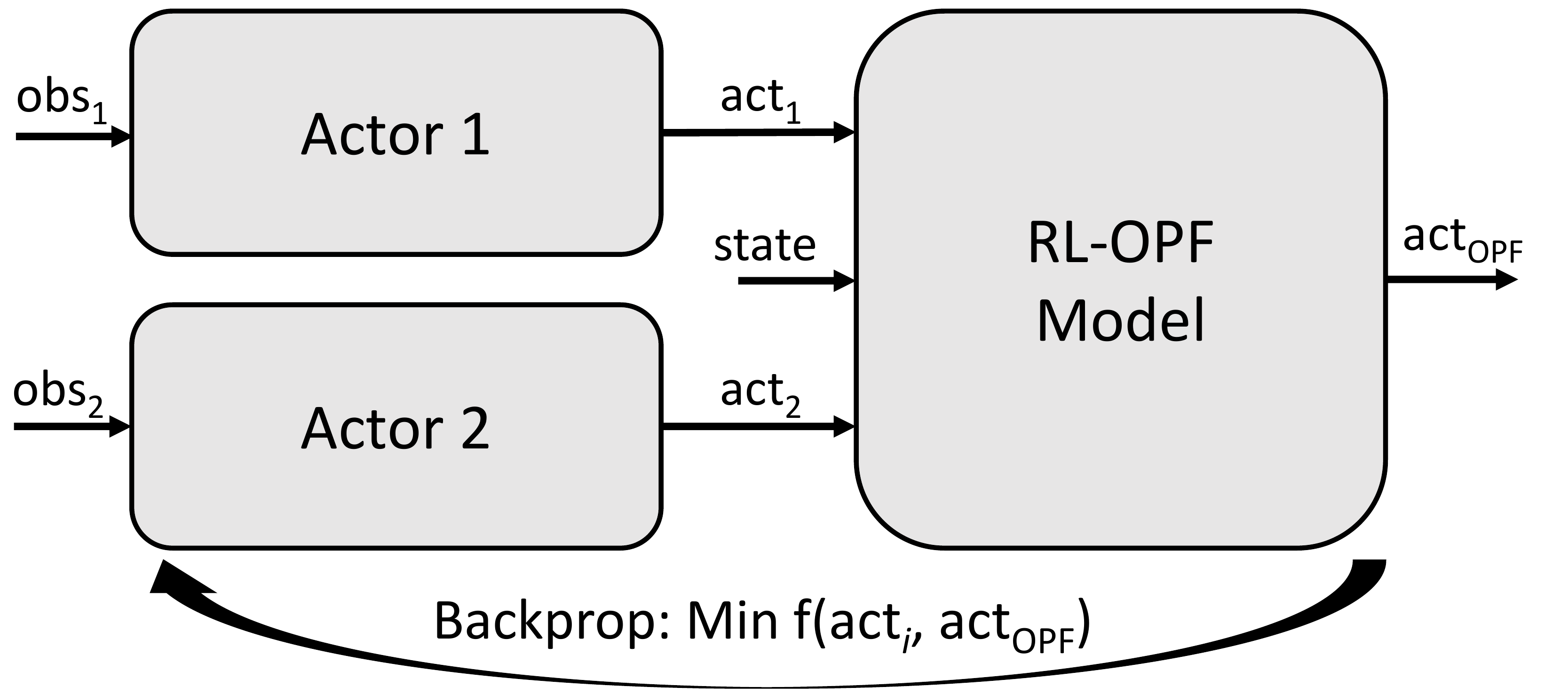}

\caption{Example of the \ac{MMADDPG} algorithm with two agents.}
\label{fig:mmaddpg}
\end{figure} 

\subsubsection{Concept 1: Learn surrogate model}\label{sec:idea1}
The first idea is to replace the \ac{OPF} for market clearing with a learned surrogate model. In section~\ref{sec:rl-opf}, we discussed the emerging body of literature on how \ac{ML} can be used to approximate the \ac{OPF}. As discussed before, the primary motivation is to speed up \ac{OPF} solving. That is especially helpful when thousands and millions of different \acp{OPF} need to be solved for the same power system, as in our scenario. Note that this applies not only to the \ac{OPF} but every expensive calculation. However, the \ac{OPF} is especially important in power system research and can serve as an example. 
\par 
The \ac{OPF} problem can be translated into an \ac{RL} problem by defining action space, observation space, and reward function. We will discuss this later in section~\ref{sec:implement1} when we apply all three concepts to a specific scenario.
\par 
Note that we can replace the environment with the surrogate in two ways. Option one is to train the surrogate to approximate the environment first and start the \ac{MARL} training only when the surrogate is finished. However, the disadvantage is that the distribution of agent bids is yet unknown. Therefore, some assumption about the distribution has to be made, \eg, uniform distribution, which may negatively affect the accuracy of the surrogate. Option two tackles that problem by training the surrogate and the \ac{MARL} algorithm in parallel. This way, the \ac{MARL} agents will improve their bidding over time, which results in better training data for the surrogate.
In this work, we apply option two to train \ac{MARL} agents and the surrogate model in parallel. However, for some scenarios, option one can be preferable as well, especially since it requires less computation.
\par 
In summary, Concept 1 is to provide the \ac{MARL} training algorithm with a learned model of the \ac{OPF} and, therefore, the market clearing. Note that any \ac{RL} algorithm can be chosen for that approach, which makes the approach modular.
\subsubsection{Concept 2: Hardcode market rules}\label{sec:idea2}
The second idea is to explicitly provide the reward function to the learning algorithm in the form of a loss function. Normally, most \ac{RL} algorithms learn to predict the future reward first to then improve their policy to maximize expected reward, \eg, \ac{DDPG} uses the expected Q-value as negative loss for the actor. However, when the reward function is known, we can explicitly provide it to the agent, which results in faster training and potentially better performance, because the loss does not suffer from approximation errors anymore.
\par 
As a simple example, one common objective of the \ac{OPF} is to minimize costs $C$, which is the product of the power setpoint $P$ and the respective price $p$.
\begin{equation}
    \mathrm{min} \; C = p \cdot P
\end{equation}
Translated to \ac{RL}, the agent wants to minimize the product of an observation -- the price -- with an action -- the setpoint. Both are known to the agent, which removes the necessity to approximate $Q$, because the cost function can be used directly as loss function for policy training. 
\par 
This concept is especially helpful in 1-step environments, which terminate after one step. In such environments, the reward is equal to $Q$. Therefore, the agent has perfect knowledge of $Q$, if it knows the reward function. As Zhen \etal \cite{hon21} showed, the \ac{OPF} approximation can be implemented as a 1-step environment because the solution of one \ac{OPF} is independent of the solution of the previous \ac{OPF}. Exceptions are multi-step \ac{OPF} problems where the optimization is done over multiple time steps, \eg, when storage systems are part of the optimization. 
\par
While especially useful for 1-step environments, the trick can also be applied to the general case, because with the current reward $r_\tau$ of step $\tau$ at least part of $Q_\tau$ is known and does not need to be learned anymore:
\begin{equation}
    Q_\tau = r_\tau^\mathrm{known} + Q_{\tau+1}
\end{equation}
The same principle is applicable when only part of the reward function is known. For example, in the \ac{OPF}, the cost minimization can be directly used as part of the loss function, but penalties for constraint satisfaction usually cannot since they are based on the yet unknown next state $s_{\tau+1}$. Again, only part of the reward function needs to be approximated, simplifying the training. 
\begin{equation}
    r_\tau = r_\tau^\mathrm{known} + r_\tau^\mathrm{unknown}
\end{equation}
%
\subsubsection{Concept 3: Backpropagate through the surrogate}\label{sec:idea3}
Previously, we discussed how learning a surrogate model of the \ac{OPF} and hardcoding the market rules into the loss function can improve training. Concept~3 builds upon both ideas by utilizing the differentiability of \acp{NN}. As a simple example, agent~$\agentidx$ wants to maximize its profit $G_{\agentidx}$, which is the product of the power setpoint $P_{\agentidx}$ of its generator and the respective price $\price_{\agentidx}$ minus some internal marginal costs $\marginal_{\agentidx}$.
\begin{equation}
    G = ( \price_{\agentidx} - \marginal_{\agentidx} )  \cdot P_{\agentidx}
\end{equation}
Normally, we cannot use hardcoding of the loss function here, because the setpoint $P_{\agentidx}$ is calculated within the environment and unknown to the agent. However, $P_{\agentidx}$ is the output of the \ac{RL}-\ac{OPF} surrogate \ac{NN}, which we provide to the \ac{MARL} learning algorithm. This way, Concept~2 is applicable again, because we can simply use $-G$ as loss function for the agent and backpropagate through the surrogate to calculate the gradients for the agent actors.
\par 
Note that the above example is for pay-as-bid pricing, where the price is equal to the bid and, therefore, the action of the agent. However, the general idea is also applicable to other schemes like uniform pricing or locational marginal pricing. In both cases, the market determines the resulting price, which makes the price an output of the surrogate model. This way, again, backpropagation through the surrogate is possible and beneficial.
\par 
In summary, instead of naively replacing the normal environment with the faster surrogate, we also utilize the differentiability of the surrogate \ac{NN}. This way, the surrogate \ac{NN} essentially serves as a central critic to all agent actors. 
The advantage is that the agents no longer need to learn a critic because the surrogate model is trained anyway. 
The resulting algorithm is visualized in Figure~\ref{fig:mmaddpg}.
%

\section{Scenario and Environment}\label{sec:scenario}
We presented three concepts on how to improve \ac{MARL} bidding in energy markets on a conceptual level. However, the exact implementation of these ideas depends heavily on the respective case, \eg, the market rules or the \ac{OPF} details. Therefore, we now present an exemplary energy market scenario, which also serves as \ac{RL} environment for our experiments later on.  
\subsection{Energy Market Bidding Scenario}
We consider an energy market bidding scenario where multiple agents operate generators and offer active power on the market.
The market clearing is done with an \ac{OPF} to prevent constraint violations in the power system. 
For simplicity, we consider a pay-as-bid market, \ie, every market participant gets paid according to their own bid. Regarding \ac{MARL}, pay-as-bid has the advantage that the agents' profits always correlate with their bids, which would not be the case for \eg uniform pricing. 
The objective function of the \ac{OPF} is to minimize total active power costs, subject to grid constraints, \ie, slack bus power flow, voltage constraints $\voltage_\mathrm{min}/\voltage_\mathrm{max}$ of buses $\buses$, line loading $S_{\branch,\mathrm{max}}$ of lines $\branches$, trafo loading $S_{\trafo,\mathrm{max}}$ of trafos $\trafos$, maximum generator power $\power^\mathrm{max}$ of agents $\agents$, and the AC nodal power balance equations (omitted here for brevity).
\begin{equation}
    \mathrm{min} \; J = \sum_{\genidx \in \agents} \power_\genidx \cdot \price_\genidx + \mathrm{max} (\power_\slack \cdot \price_\mathrm{max}, 0)
\end{equation}
\begin{align}\label{eq:constraints}
    \mathrm{s.t.} \quad 
     & \voltage_\mathrm{min} \leq \voltage_\bus \leq \voltage_\mathrm{max} \; \forall \; \bus \in \; \buses \\
     & S_\branch \leq S_{\branch,\mathrm{max}} \; \forall \; \branch \in \; \branches \\
     & S_\trafo \leq S_{\trafo,\mathrm{max}} \; \forall \; \trafo \in \; \trafos \\
     & 0 \leq P_\agentidx \leq \power_\agentidx^\mathrm{max}  \; \forall \; \agentidx \in \; \agents
\end{align}
Each agent $\agentidx$ operates a single generator and provides active power $\power_\agentidx$ for price $\price_\agentidx$. 
Also note that the power flow from the slack bus $\power_\slack$ is implemented as a soft-constraint with a penalty $\price_\mathrm{max}$ to always have a valid solution within the constraints.
\par 
In our environment, we use the open-source tool \pandapower\pandapowerfootnote \cite{thu18} for the \ac{OPF} calculation and the \simbench\simbenchfootnote \cite{mei20} benchmark system \texttt{1-HV-urban--0-sw} with 372 buses and 42 generators as a power model. To generate realistic power system states, we use the associated full-year time-series data of the system. Further, we add noise of $\pm 10\%$ to the loads to prevent repetition of the quarter-hourly data samples and to increase variance. 
\par 
To investigate a potential influence of the number of agents $\nagents$, the environment should allow for a flexible number of agents, and therefore generators. However, to prevent an unrealistic setting, we consider the total generator capacity of $\power_\total$ of the system as constant and evenly distribute it to all agents \agents. The resulting active power capacity $\power_\agentidx^\mathrm{max}$ of each agent/generator is:
\begin{equation}
    \power_\agentidx^\mathrm{max} = \dfrac{\power_\total}{\nagents} \; \forall \;  \agentidx \; \in \; \agents
\end{equation}
The locations of the generators are randomly selected from the 42 generator locations in the original \simbench system.


\subsection{Observations, Actions, and Rewards}
We model the energy market bidding problem as a 1-step partially observable Markov game. We assume that the market participants have zero knowledge about the system's physical state and observe only the current time $\tau$, which results in partial observability. To provide time as observation, we encode it as sine/cosine pairs for day-time, week-time, and year-time, which makes six observations. The sine/cosine encoding has the advantage that 24:00 and 00:00 are identical in the observation space instead of being at maximum distance to each other.
\begin{equation}
    \obs_{1,2,3} = \sin \! \left( 2\pi\cdot \frac{\tau \; \% \; \mathit{tf}}{\mathit{tf}} \right) \; \forall \; \mathit{tf} \in \mathit{TF}
\end{equation}
\begin{equation}
    \obs_{4,5,6} = \cos \! \left( 2\pi\cdot \frac{\tau \; \% \; \mathit{tf}}{\mathit{tf}} \right) \; \forall \; \mathit{tf} \in \mathit{TF}
\end{equation}
With \% as modulo function and the three time-frames
\begin{equation}
    \mathit{TF} = \{ 4 \cdot 24, \; 4 \cdot 24 \cdot 7, \; 4 \cdot 24 \cdot 366 \}
\end{equation}
for day, week, and year respectively.
For simplicity, we define only the bids of the agents as actions. Every agent has a single continuous action in the range $\pricerange$. We assume that all agents always offer all their active power capacity, without withholding capacity from the market.
\begin{equation}
    \act \in \pricerange
\end{equation}
The reward of each agent $\agentidx$ is its profit on the market, \ie, the product of the price $\price$ minus some marginal costs $\marginal$ and the resulting active power setpoint $\power$.
\begin{equation}\label{eq:reward}
    r_\agentidx = ( \price_\agentidx - \marginal_\agentidx)  \cdot \power_\agentidx
\end{equation}
The active power setpoints are determined by the market clearing, \ie, the \ac{OPF}. The marginal costs are assumed to be constant and to be 10\% of the maximum price $\price_\mathrm{max}$, which is chosen as 600~€/MW.
%
%
\section{Implementation of the Concepts}\label{sec:implement}
In this section, we discuss how we applied each conceptual idea of the \ac{MMADDPG} algorithm to this specific scenario. 

\subsection{Concept 1: Learn Surrogate Model}\label{sec:implement1}

To replace the \ac{OPF} with a learned surrogate model, we approximate it with another \ac{DRL} algorithm. For that, we model the \ac{OPF} problem as a 1-step \ac{MDP} \cite{zhe21}. To train this \ac{OPF}-agent that represents the market, again, action space, observation space, reward, and \ac{RL} algorithm need to be chosen.
\par 
In the case of market clearing of an energy market the actions of the \ac{RL} agent are the active power setpoints $\power_\agentidx$ of all generators in the system.
\begin{equation}
    \act_\mathrm{OPF} = \power_\agentidx \in [0, \power^\mathrm{max}_\agentidx] \; \forall \; \agentidx \in \agents
\end{equation}
\par 
We assume that the state of the power system is fully known to the \ac{OPF}-agent, except for the yet unknown actions, \ie, the generator setpoints. Under the assumption of a fixed network topology, active and reactive power of all loads $\loads$ in the system are sufficient. Additionally, the \ac{OPF}-agent observes the active power prices of all generators, \ie, their bids on the market.
\begin{equation}
    \obs_\mathrm{OPF} = \{\power_l, Q_l, \price_\agentidx\} \; \forall \; \loadidx \in \loads \; \mathrm{and} \; \forall \; \agentidx \in \agents 
\end{equation}
\par 
We define the reward function as the negative objective function $J$ of the \ac{OPF} minus linear penalties $\penaltyOPF$ for each constraint. 
\begin{equation}
    r_\mathrm{OPF} = -J - \penaltyOPF_\mathrm{voltage} - \penaltyOPF_\mathrm{line} - \penaltyOPF_\mathrm{trafo}
\end{equation}
with penalties $\penaltyOPF$ for the constraints (compare eq. (\ref{eq:constraints}) ff.):
\begin{equation}
    \penaltyOPF_\mathrm{voltage} = \sum_{\bus \in \buses} \mathrm{max}( \voltage_\bus - \voltage_\mathrm{max}, \voltage_\mathrm{min} - \voltage_\bus, 0)
\end{equation}
\begin{equation}
    \penaltyOPF_\mathrm{line} = \sum_{\branch \in \branches} \mathrm{max} \! \left( \frac{S_\branch}{S_\mathrm{max}} - 100\%, 0 \right)
\end{equation}
\begin{equation}
    \penaltyOPF_\mathrm{trafo} = \sum_{\trafo \in \trafos} \mathrm{max} \! \left( \frac{S_\trafo}{S_\mathrm{max}} - 100\%, 0 \right)
\end{equation}
Note that power flow balance constraints are automatically met when a powerflow calculation is done in the environment to compute the next system state. 
\par 
Any \ac{DRL} algorithm for continuous action spaces could be applied to the previously defined \ac{RL}-\ac{OPF} task. In this work, we use \ac{DDPG} to learn the \ac{OPF}, mainly because of two advantages: As an off-policy algorithm, \ac{DDPG} is very sample-efficient. That is important since the computation of the grid state still requires a powerflow calculation, which is computationally heavier than most benchmark \ac{RL} environments--although far less demanding than the actual \ac{OPF}. The second advantage is that the utilized \ac{DDPG} and \ac{MMADDPG} can share their replay buffer for training data.
%
\subsection{Concept 2: Hardcode Market Rules}\label{sec:implement2}

In the previous section, we discussed how part of the environment can be replaced with a learned \ac{OPF} model, which we provide to \ac{MMADDPG} to speed up multi-agent learning. 
With the \ac{RL}-\ac{OPF} surrogate model, \ac{MMADDPG} now has access to all parts of the agents' reward function (\ref{eq:reward}). The bid is the agent's own action, the marginal costs are constant, and the active power setpoint is the output of the \ac{RL}-\ac{OPF}. Therefore, we can directly hardcode the actor loss $l_\agentidx^\mathrm{actor}$ of agent $\agentidx$ as:
\begin{equation}
    l_\agentidx^\mathrm{actor} = -r_\agentidx
\end{equation}
Note that if the marginal costs were not constant, they would be required to be part of the agent's observation space. 
\par 
The hardcoding of the market rules and the profit reduces training effort and removes one source of approximation error in our learned model of the market. However, it is important to remember that the \ac{RL}-\ac{OPF} model is still only an approximation, which results in a non-perfect gradient signal for the actors, similar to a normal critic network. 
\par 
The trick of hardcoding the market rules cannot only be done for \ac{MADDPG} but also for the \ac{DDPG} algorithm that approximates the \ac{OPF}. The goal of the \ac{OPF} is to minimize costs on the market while adhering to all constraints. The resulting loss function can be written as:
\begin{equation}
    l_\mathrm{OPF}^\mathrm{actor} = \sum_{\genidx=1}^{\agents} \power_\genidx \cdot \price_\genidx - Q^\mathrm{penalties}
\end{equation}
In contrast to the agents' loss, not the entire function is known here because voltages and line loads would be required for penalty prediction, but are unknown. Therefore, \ac{DDPG} here still requires a critic but only needs to learn the part of the Q-function that represents the constraints, as discussed in section~\ref{sec:idea2}. 
\subsection{Concept 3: Backpropagate Through Surrogate}\label{sec:implement3}
The first part of Concept 2 is only possible because the \ac{OPF} is performed by a neural network, which makes it fully differentiable. 
This way, the active power setpoints $\power_\agentidx$ can be used as part of the loss function by using backpropagation. The agents can optimize their bidding behavior to maximize profit under consideration of the market rules, \ie, the \ac{OPF}, without any need to learn the market rules. No additional implementation is required here because of pytorch's automatic differentiation package \texttt{autograd}.
\section{Experimentation}\label{sec:experimentation}
In the following, we will discuss the hyperparameter settings for training and introduce regret as a distance metric for measuring \ac{NE} approximation. 
\subsection{Hyperparameters}\label{sec:hyperparams}
The chosen hyperparameters for all three utilized \ac{RL} algorithms are listed in Table \ref{tab:hyperparams}.
\begin{table}[ht]

\centering
\caption{Hyperparameters of the three DRL algorithms.}
\footnotesize
\begin{tabular}{l|l|l|l}
Hyperparameter          & \ac{MADDPG}  & M-\ac{MADDPG} & \ac{DDPG}  \\ \hline
batch size              & 256           & 256       & 128 \\
actor learning rate     & 0.001         & 0.001     & 0.0001 \\
critic learning rate    & 0.001         & //        & 0.001 \\
actor neurons/layer    & (128,)           & (128,)       & (256,) \\
critic neurons/layer   & (256,)           & //        & (256,) \\
optimizer               & RMSprop    & RMSprop   & Adam   \\
noise std               & 0.2           & 0.2       & 0.2 \\
start train             & 1000          & see eq. (\ref{eq:starttrain}) & 1000   
\label{tab:hyperparams}
\end{tabular}
\end{table}
When applicable, we chose the same hyperparameters for \ac{MADDPG} and \ac{MMADDPG}. 
\par 
The agent training of \ac{MMADDPG} starts when the internal \ac{RL}-OPF is already trained quite well because otherwise the \ac{OPF} model is useless for gradient computation. However, since the \ac{OPF} approximation gets slower with rising number of agents/generators, we increase the start of the agents' training with the number of agents $\nagents$.
\begin{equation}\label{eq:starttrain}
    \mathrm{start \; train} = \mathrm{max} (150 \cdot \nagents, 2000)
\end{equation}
\par
Note that we utilize the RMSprop optimizer 
for both \ac{MARL} algorithms instead of the often-used Adam because we observed a significant performance improvement in both cases. The momentum-based Adam 
is probably not well suited for \ac{MARL} since the momentum is derived from older data, where the competing agents had different behavior and therefore are outdated. However, the exact reason is out-of-scope here.  
\par 
\subsection{Testing and Metrics}\label{sec:regret_test}
In a competitive \ac{MARL} problem, the agents' reward is not suitable to measure the training success. For example, in a market environment, decreasing rewards (profits) can be expected during training, because agents must underbid each other to make any profit.
Therefore, we utilize regret as a metric to measure the \ac{MARL} algorithm's success in learning bidding strategies \cite{yan20c}.
The regret $\regret$ is defined as the maximum possible reward  $\bestreward$ minus the actual reward $\reward$:
\begin{equation}\label{eq:regret}
    \regret = \bestreward - \reward
\end{equation}
If the total regret of all agents is zero, no agent has an incentive to change strategies, which defines a \ac{NE} \cite{yan20c}. However, regret can also serve as a distance metric for how well the equilibrium point was approximated and how much the agents would want to change their respective strategies. That enables the comparison of algorithms regarding their performance to approximate expected bidding behavior in market scenarios.
\par 
Usually, the optimal reward required to compute the regret is unknown. However, since all agents have only one action, we can apply a simple heuristic: First, we store the current bids of all agents $\agents$ after training. Second, for agent $\agentidx$, we sample some equidistant bids in the full bid range $\pricerange$ and calculate the agent's profit with the \ac{OPF} market clearing. Third, we iteratively perform a local search around the current best bid until convergence. Fourth, we calculate the regret for agent $\agentidx$ with equation (\ref{eq:regret}). We repeat steps two to four for each agent and calculate the total regret $\totalregret$:
\begin{equation}
    \totalregret = \sum_{\agentidx \in \agents} \regret_\agentidx
\end{equation}
To account for the variety of different system states, we perform this test \ntest~times for different states and compute the average for evaluation of the training. %

\section{Results}\label{sec:results}
To evaluate the performance of the proposed concepts, we apply \ac{MMADDPG} and basic \ac{MADDPG} to the \ac{MARL} bidding scenario presented in section~\ref{sec:scenario}. To investigate the influence of the number of agents, we apply both algorithms to variations of the same scenario with 10, 20, 30, and 40 agents, with 30 and 40 agents being higher than the previous state of the art. We repeat each training run $\nseed$ times to compensate for the stochasticity of \ac{RL} training \cite{hen18}. 
All experiments are done on a DGX-1 deep learning server.
\par
\begin{figure}[t]
    \centering
	\includegraphics[width=80mm]{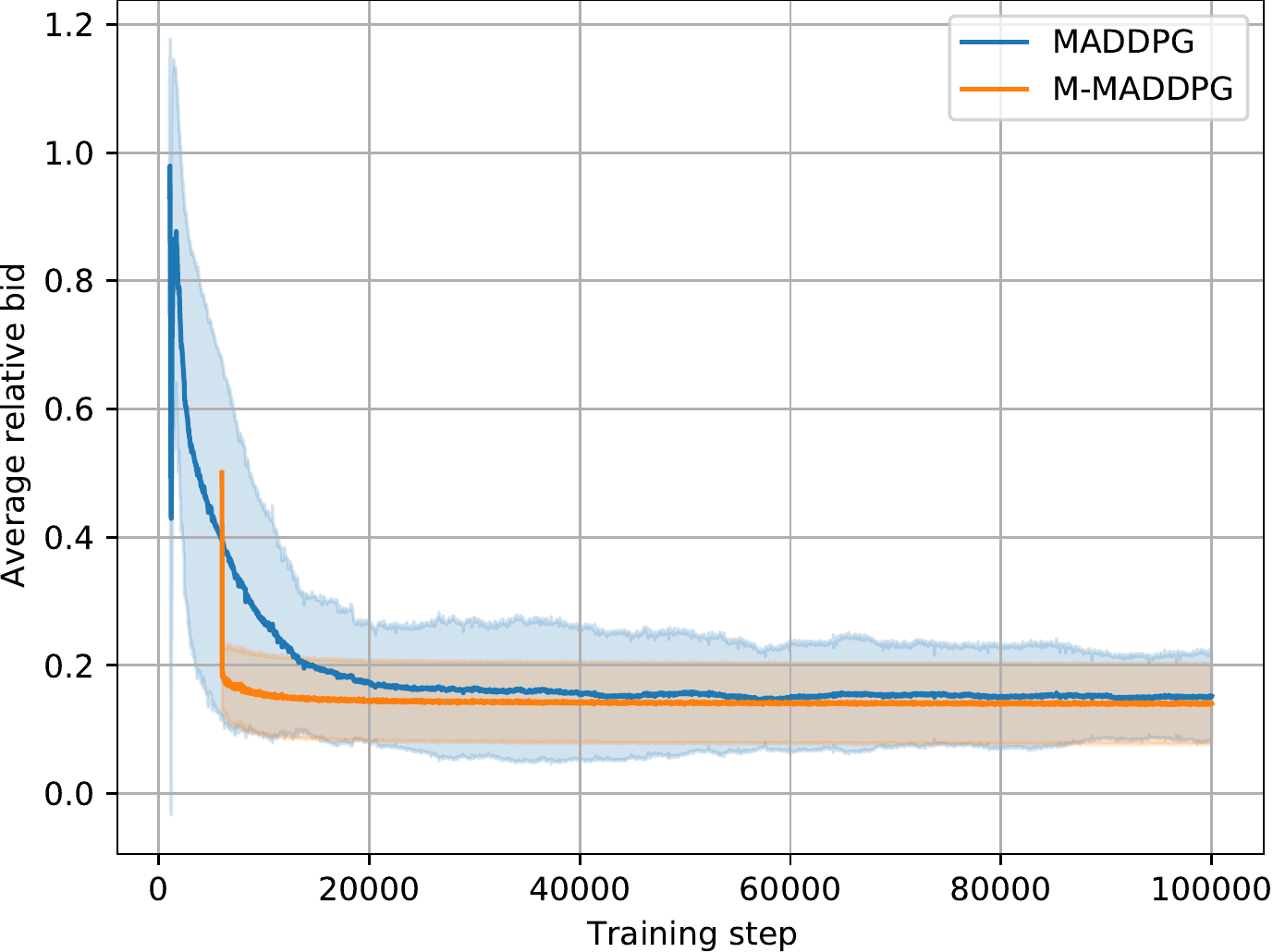} 
	\caption{Average bids and standard deviation of MADDPG and M-MADDPG agents over 100k training steps, averaged over 40 agents and $\nseed$ runs.}
	\label{fig:bidding}
\end{figure}

 
We compare the resulting bidding behavior of both algorithms, their capability to minimize regret, the influence of the number of agents, and computation time until convergence.
Fig.~\ref{fig:bidding} shows the resulting average bidding behavior with both algorithms relative to the maximal bid $\price_\mathrm{max}$ by the example of the 40-agent case. 
We can observe two things: First, for both algorithms, the agents learn to bid slightly above their marginal costs of 0.1. Second, \ac{MMADDPG} converges significantly faster, smoother, and also to slightly lower average bids. Especially at the beginning, the agents immediately jump to bids around 0.2, which is close to the final result. Note that the training of \ac{MMADDPG} starts delayed, as discussed in section~\ref{sec:experimentation}. 
\par
\begin{figure}[t]
    \centering
	\includegraphics[width=80mm]{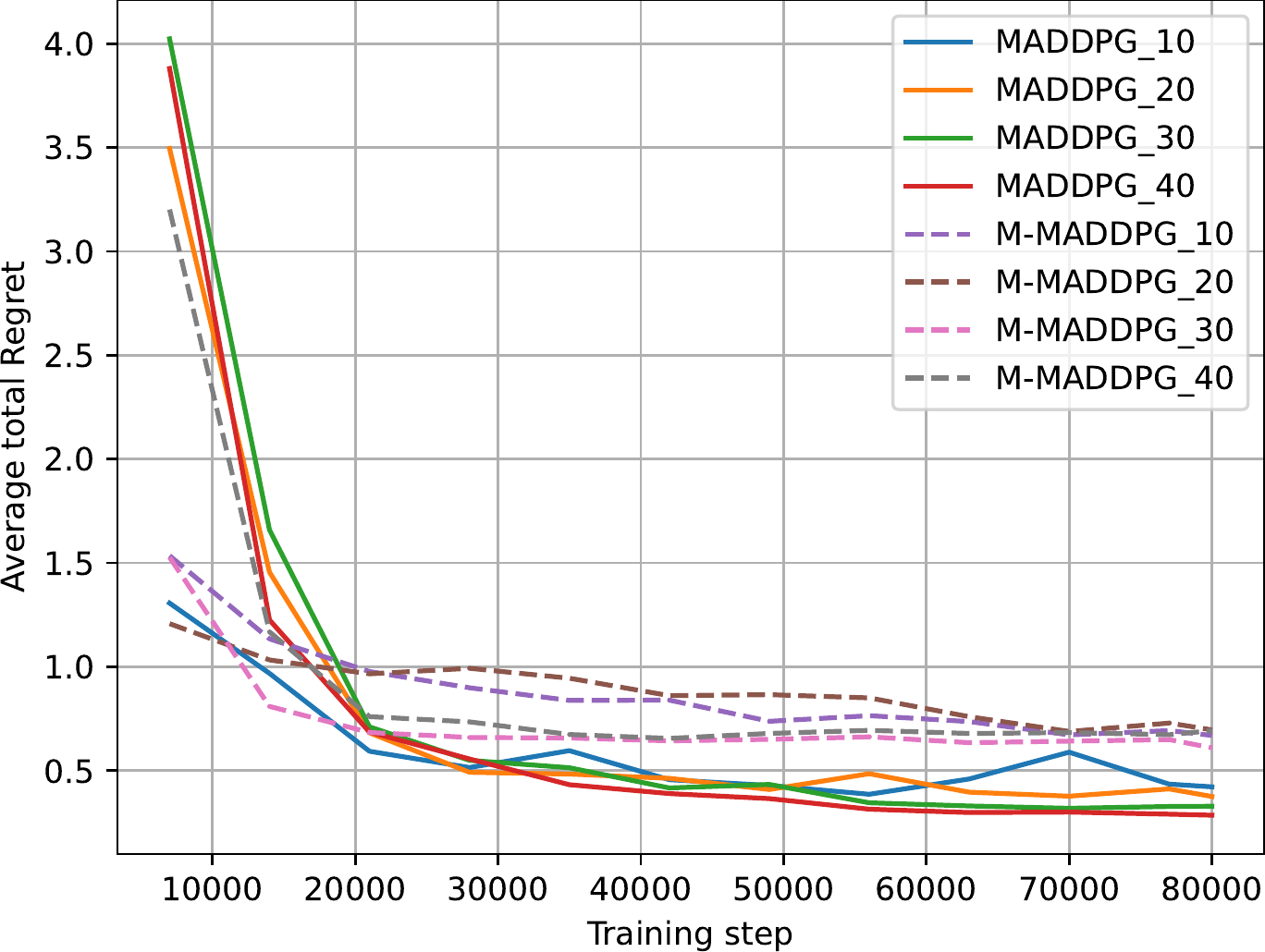}
	\caption{Total regret of MADDPG and M-MADDPG over the training course. Averaged over 3 runs respectively.}
	\label{fig:regret}
\end{figure}

 
Fig.~\ref{fig:regret} visualizes the regret course over training for both algorithms for 10, 20, 30, and 40 agents. For this, we stopped the experiment regularly to perform the regret test described in section \ref{sec:experimentation}. Because the tests require many computationally expensive \ac{OPF} calculations, the results are averaged over three runs this time.  
\par 
Fig.~\ref{fig:regret} shows that both algorithms can minimize the total regret in all cases. Second, \ac{MADDPG} results in significantly lower final regret in all four cases. Third, the model-based \ac{MMADDPG} results in faster convergence again, and therefore lower regrets until about time step 20k, except for the case with 10 agents, where \ac{MADDPG} is equally fast. Fourth, the convergence of \ac{MADDPG} slows down the more agents are considered. For \ac{MMADDPG}, the opposite is true; more agents result in faster convergence.
\par
\begin{figure}[t]
    \centering
 	\includegraphics[width=80mm]{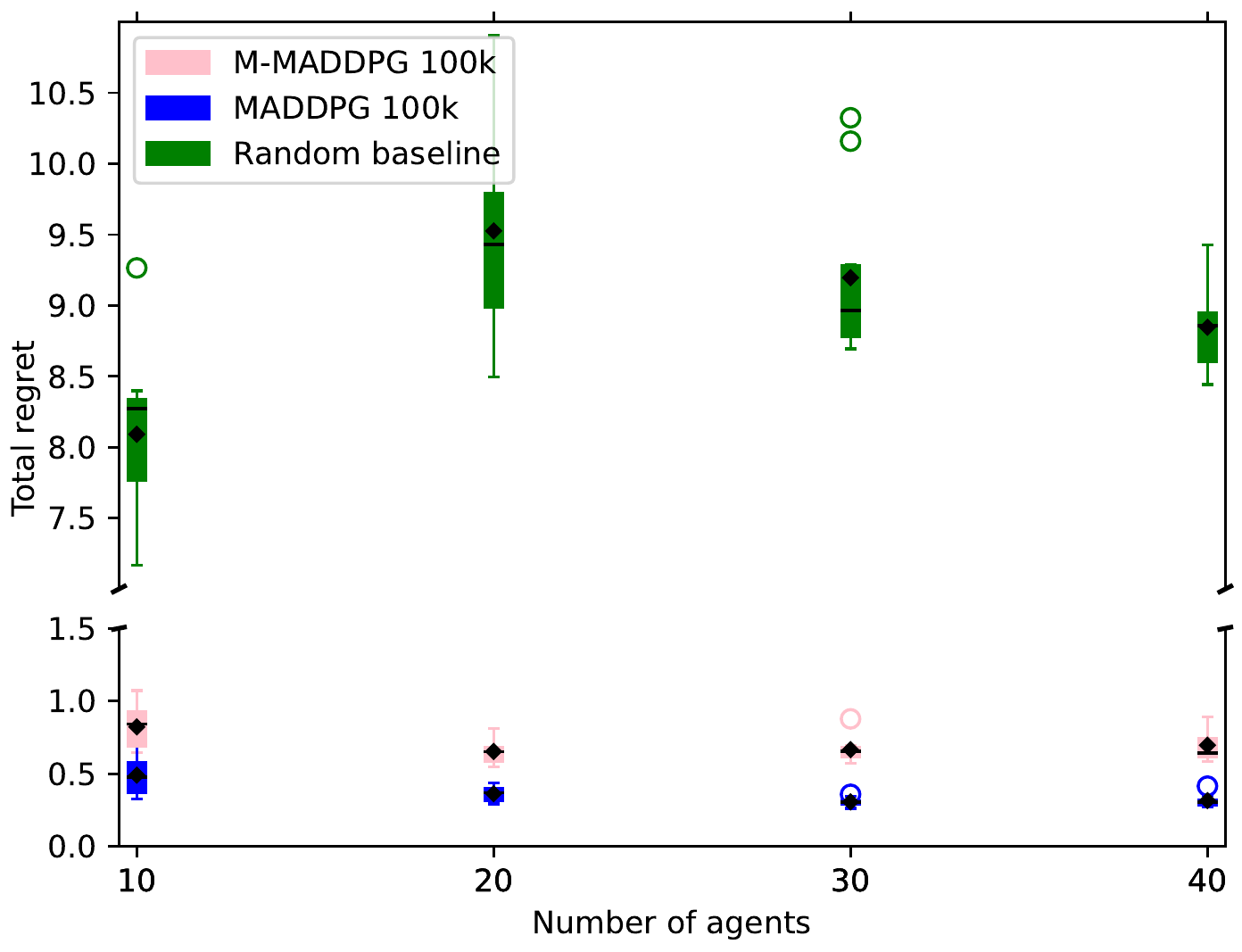} 
	\caption{Total regret after 100k training steps for \ac{MADDPG}, \ac{MMADDPG}, and random bidding behavior for 10, 20, 30, 40 agents. Averaged over 10 runs respectively.}
	\label{fig:comparison100}
\end{figure}

 
Fig.~\ref{fig:comparison100} shows the final regret distribution after training for 100k steps of both algorithms for 10, 20, 30, and 40 agents with boxplots. As a baseline, we also visualize the regret distribution if all agents acted randomly.
Again, \ac{MADDPG} results in slightly lower regrets in all four cases. However, both algorithms significantly outperform the random baseline by a factor of about 10 to 20. Notice the broken y-scale to visualize the random baseline. 
Finally, we observe that the regret for both algorithms is highest in the 10-agent case.
\par
\begin{figure}[t]
    \centering
	\includegraphics[width=80mm]{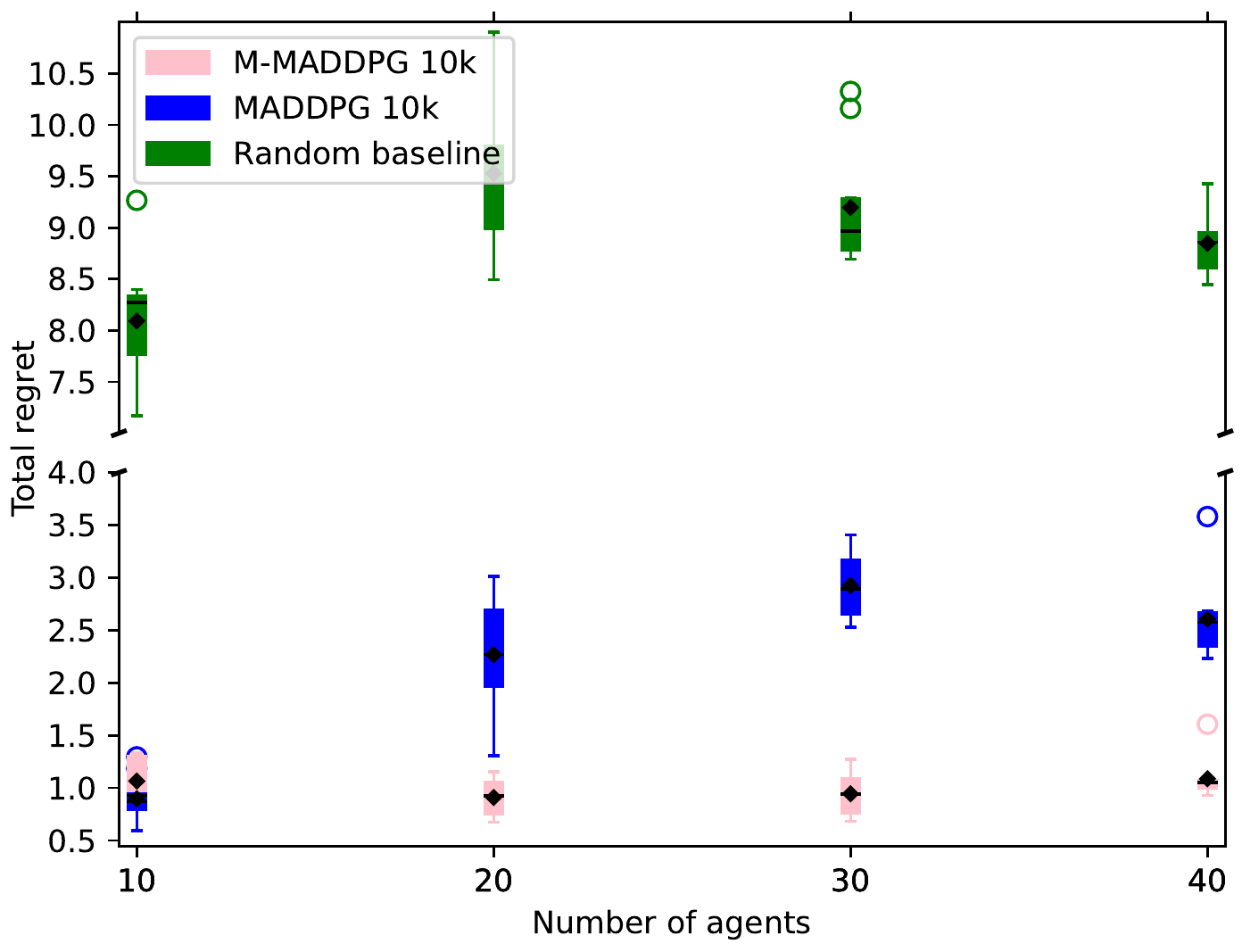} 
	\caption{Total regret after 10k training steps for \ac{MADDPG}, \ac{MMADDPG}, and random bidding behavior for 10, 20, 30, 40 agents. Averaged over 10 runs respectively.}
	\label{fig:comparison10}
\end{figure}

 
Fig.~\ref{fig:regret} indicated faster convergence of \ac{MMADDPG}. Therefore, Fig.~\ref{fig:comparison10} shows the final regret after only 10k training steps. 
With shorter training, \ac{MMADDPG} consistently outperforms \ac{MADDPG} by a factor of two to three. The exception is the 10-agent case again, where both algorithms result in roughly the same regret distribution. Even with shorter training, both algorithms significantly outperform the random baseline by a factor of three to nine.
\par 
One intended benefit of \ac{MMADDPG} was less computation time per training step. 
Tab.~\ref{tab:speedup} shows the average training times of \ac{MADDPG} to \ac{MMADDPG} for 100k training steps. 
%
\begin{table}[t]
\centering
\caption{Average training time of \ac{MMADDPG} and \ac{MADDPG} for 100k training steps for 10, 20, 30, and 40 agents.}
\footnotesize
\begin{tabular}{l|l|l|l|l}
\multirow{2}{*}{Algorithm}                 & \multicolumn{4}{l}{Train time} \\
          & 10 agents & 20 agents & 30 agents & 40 agents  \\ \hline
\ac{MADDPG}    & 73.5 h & 89.74 h & 111.92 h & 143.74 h \\
\ac{MMADDPG}   & 7.5 h & 13.2 h & 21.66 h & 34.58 h \\
Speed-up ratio       & 9.81 & 6.53 & 5.17 & 4.16
\label{tab:speedup}
\end{tabular}
\end{table}
%
The average training time is in the range of one to multiple days and increases together with the number of agents. \ac{MMADDPG} learns significantly faster than \ac{MADDPG} in all cases, from four times faster for 40 agents to ten times faster for 10 agents. 
Considering that \ac{MMADDPG} also converged faster in all cases, the actual speed-up would be even higher if we interrupted training earlier, especially for high agent numbers. For example, the 40 agent runs in Fig.~\ref{fig:regret} converged at around 20k for \ac{MMADDPG} and 60k steps for \ac{MADDPG}, resulting in a total speed-up factor of roughly $3 \cdot 4 = 12$ on average.

\par
\begin{figure}[t]
    \centering
	\includegraphics[width=80mm]{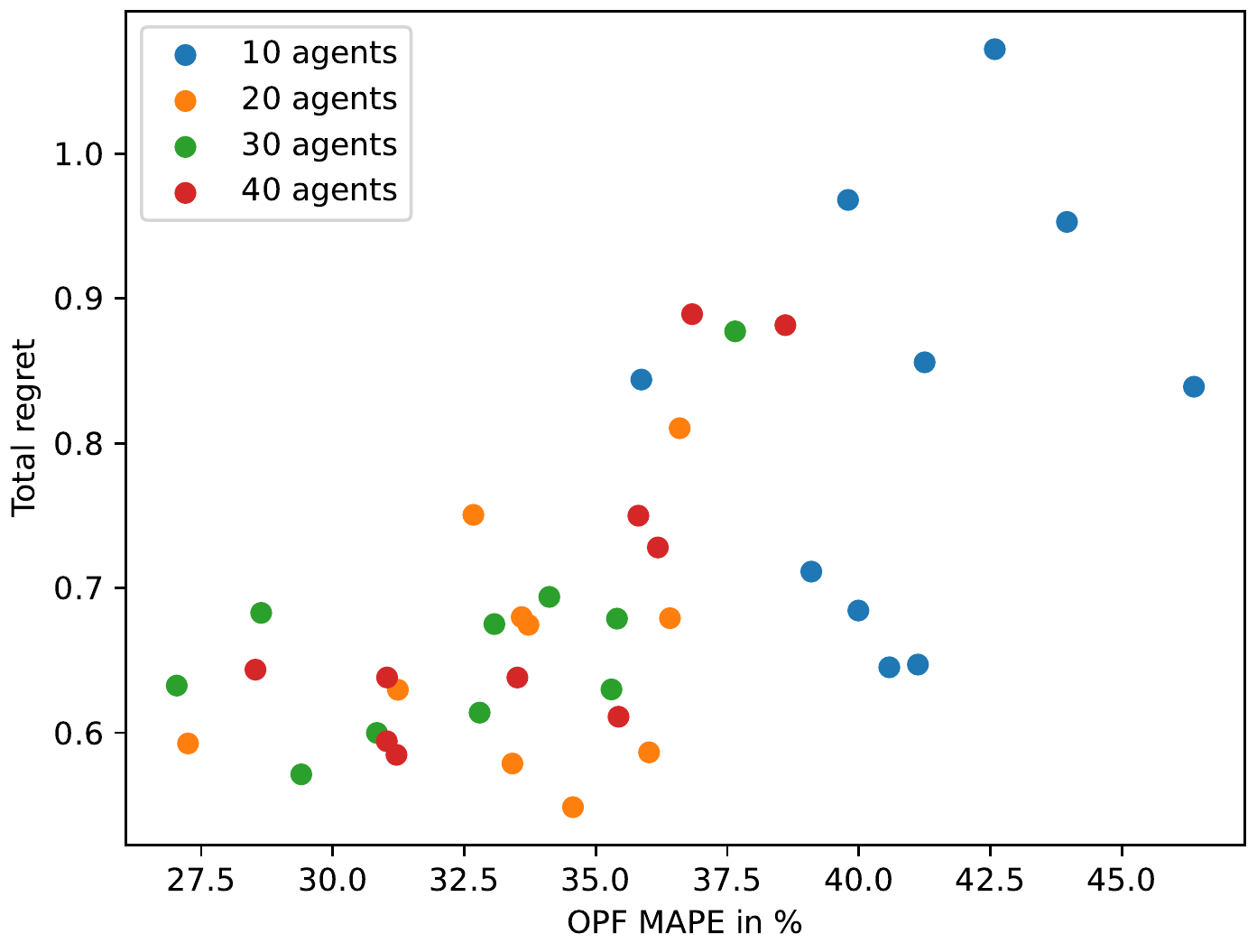} 
	\caption{Total regret after 100k training steps for \ac{MMADDPG} in relationship to the MAPE of the \ac{OPF} approximation at the end of training. 40 samples are shown; 10 for each scenario with 10, 20, 30, and 40 agents respectively.}
	\label{fig:opf_correlation}
\end{figure}

 
The speed-up in training time was achieved by adding a model to the \ac{MADDPG} algorithm. That model was built by approximating the \ac{OPF} for market clearing with \ac{RL}. Since this approximation is not perfect, we expect a correlation between the quality of \ac{OPF} approximation and the resulting total regret. Fig.~\ref{fig:opf_correlation} shows a scatter plot of all 40 runs with \ac{MMADDPG} and 100k training steps.
The \ac{MAPE} of the \ac{OPF} approximation was computed by comparing the costs of the \ac{RL}-\ac{OPF} with the \ac{OPF} solution from \pandapower, averaged over 500 random data samples respectively. It lies in the range of about 27\% to 46\%.  
The data indicate a moderate to strong correlation between the \ac{MAPE} of the \ac{OPF} approximation and the resulting regret. The Spearman correlation is 0.67 with a p-value of $2.77 \cdot 10^{-6}<0.05$, showing clear statistical significance. 


\section{Discussion}\label{sec:discussion}
Fig.~\ref{fig:bidding} and \ref{fig:regret} demonstrate that both algorithms, \ac{MADDPG} and model-based \ac{MMADDPG}, lead to expected behavior in a competitive market situation. With both algorithms, the agents learn to bid lower and lower to underbid their competitors. And since all agents do this, the average bidding converges to a value slightly above their marginal costs, which is expected behavior in a pay-as-bid setting with imperfect competition \cite{hag12}. 
This also explains why the regret decreases together with the bidding. Lower bidding of the agents decreases their potential for profit, even if their bid is accepted, and therefore reduces the regret as well. Again, this happens for all agents in parallel due to competition. 
We can conclude that both algorithms learn meaningful and expected market behavior. In both cases, the regret is reduced drastically below the random baseline, indicating a convergence to some \ac{NE}. However, the total regret did not converge to exactly zero.  
\par 
The two algorithms differ in speed of convergence and the final regret. \ac{MMADDPG} converges significantly faster and more stable regarding bidding and regret. Since the fundamental learning algorithm remains the same as \ac{MADDPG}, this can only be explained by the \ac{OPF} model and the hardcoded market rules that the algorithm has access to. While \ac{MADDPG} needs to explore first what good and bad actions are, \ac{MMADDPG} can learn meaningful behavior immediately, due to the model. The benefits of this faster learning even compensate for the delayed training start until the \ac{RL}-\ac{OPF} model is sufficiently trained.
\par 
The downside of \ac{MMADDPG} is that the final regret is worse than for \ac{MADDPG}. Again, this can be explained by the learned model of \ac{MMADDPG}. Fig.~\ref{fig:opf_correlation} shows a noteworthy error of the \ac{RL}-\ac{OPF} compared to the actual optimal solutions. Since the actor \ac{NN} of the \ac{RL}-\ac{OPF} is used directly to compute the gradients for the agent policies, this error gets backpropagated into the agent policies. Because the market model is erroneous, the resulting agent behavior is erroneous as well. This leads to a higher regret compared to \ac{MADDPG}, which was trained with the \pandapower \ac{OPF}. 
\par 
On the other hand, the learned \ac{RL}-\ac{OPF} resulted in a significant speed-up per training step. We already discussed that fewer training steps--\ie interactions with the environment--are required until convergence of \ac{MMADDPG}. In addition, every interaction with the environment is faster because only a power flow calculation needs to be done instead of an \ac{OPF} calculation. Tab.~\ref{tab:speedup} demonstrates the resulting speed-up of factor four to ten compared to \ac{MADDPG}. The speed-up is especially big for low numbers of agents because higher agent numbers shift the computational efforts more and more from environment computation to the optimization of the agent neural networks, which was not optimized in \ac{MMADDPG}. 
The combined speed-up of less training time per step and faster convergence is always around one order of magnitude. For small agent numbers, this effect is dominated by the step-wise speed-up, and for high agent number it is dominated by the faster convergence.
Note that we performed all runs on CPUs only to allow for a fair comparison. Since \ac{MMADDPG} requires additional \ac{NN} training and less extensive environment computation, we expect the speed-up to be even higher if we utilized GPUs for training because \ac{NN} training benefits from GPU usage, while the \ac{OPF} does not. 
In the following, we will summarize and discuss the benefits and drawbacks of the presented concepts, in comparison to \ac{MADDPG} and also in general.
\subsection{Benefits of M-MADDPG}
\hyperref[sec:idea1]{Concept~1} of \ac{MMADDPG} was to replace the environment with an \ac{RL}-learned surrogate model of the market, in this case, the \ac{OPF} for pay-as-bid market clearing. The main benefit is the non-need for an actual \ac{OPF} implementation. For example, if we wanted to repeat the experiments with uniform pricing, we could replace the reward definition and approximate a different \ac{RL}-\ac{OPF}. \ac{MADDPG}, however, would require a completely different implementation of the \ac{OPF}, because the utilized \pandapower \ac{OPF} cannot deal with the uniform pricing scheme.
\par 
Besides the reduced implementation time, \ac{MMADDPG} also results in faster training in two ways. First, since the \ac{OPF} calculation in the environment is replaced by a trained neural network (\hyperref[sec:idea1]{Concept~1}) the training time is reduced drastically. That effect is especially strong for low numbers of agents. 
Second, the model-based training results in faster convergence because the agents know the market rules from the start (\hyperref[sec:idea2]{Concept 2}) and because we can backpropagate through the model (\hyperref[sec:idea3]{Concept 3}). This effect is stronger for a larger number of agents. 
\par 
The next benefit is the modularity of \ac{MMADDPG}.
Since any other continuous \ac{DRL} algorithm could be used instead of \ac{DDPG}, that results in a modular algorithm, which benefits from all further advances in single-agent \ac{RL}. In future research, \ac{DDPG} can be replaced by a more advanced algorithm.
This is reinforced by Fig.~\ref{fig:opf_correlation}, which shows a correlation between the \ac{RL}-\ac{OPF} approximation and the final regret of \ac{MMADDPG}, which can be assumed to be a causal relationship. This way, \ac{MMADDPG} will automatically benefit from advances in single-agent \ac{RL} research.
\par 
Finally, the concepts presented in section \ref{sec:mmaddpg} are applicable to other use cases than the market bidding scenario presented here. Instead of learning the \ac{OPF}, any heavy computation in the environment can be approximated by neural networks to speed up training. The hardcoding of (parts of) the reward function (\hyperref[sec:idea2]{Concept 2}) is also possible in other scenarios.
%
\subsection{Drawbacks of M-MADDPG}
The main drawback of using the learned \ac{RL}-\ac{OPF} model for training the market agents is that all errors in the model get backpropagated. A non-perfect model will also result in non-optimal learning of the agents, sometimes because they receive wrong signals, sometimes because they can exploit errors in the \ac{RL}-\ac{OPF} market model. Fig. \ref{fig:comparison100} shows that the resulting regret is still in the same range as the \ac{MADDPG} results. However, the error is noteworthy and statistically significant.
\par 
\par 
Further,  \ac{MMADDPG} suffers from unsolved problems in \ac{RL}-\ac{OPF} approximation. For example, it is still unclear how to adhere to hard-constraints of the \ac{OPF} with \ac{RL} methods, which is why we used penalties as soft-constraints here, as it is state of the art, \eg, used in \cite{yuh21, woo20, yan20}. However, the soft-constraints change the properties of the \ac{OPF}, \eg, the grid operator is not forced to achieve constraint satisfaction with its actions. Therefore, the \ac{MMADDPG} agents are not able to exploit these constraints with their bidding behavior, \eg, by systematically bidding higher in high load situations.  
We assume that the non-existence of hard-constraints in the \ac{RL}-\ac{OPF} is one of the main reasons for the higher regret compared to base \ac{MADDPG}. 



\section{Conclusions}\label{sec:conclusion}
We applied \ac{MARL} to learn the market behavior of up to 40 market participants in an energy market setting. For that, we presented multiple concepts of how domain knowledge can be added to basic \ac{MARL} algorithms by the example of \ac{MADDPG}. We published our market environment together with all other source code to serve as a benchmark for further advances in the energy market bidding problem.
\par 
Our model-based approach \ac{MMADDPG} speeds up training drastically and also makes the implementation of an \ac{OPF} for training unnecessary. 
Both the basic \ac{MADDPG} and our \ac{MMADDPG} converged to meaningful market behavior and reduced the regret metric drastically compared to the baseline. 
Further, we increase the state of the art of applying \ac{MARL} to the energy market bidding problem from 25 to 40 agents.
\par
In the long term, the approach of learning market behavior with \ac{MARL} is applicable to market design, modeling of energy markets and their participants, and investigation of manipulative strategies and their respective countermeasures. 
Because of the drastic speed-up, our ideas are especially applicable if an optimization or some other heavy computation is required in the environment. For example, further applications could be re-dispatch markets or reactive power markets, which are usually cleared by an \ac{OPF} as well \cite{wol22}.
The application in market design seems especially promising. \ac{MMADDPG} is fast and does not require an explicit market implementation, which allows for rapid simulation of diverse market design variants. This way, the markets can be evaluated and compared regarding their ability to yield desired behavior of the market participants. Note that for such practical analyses, no exact \acl{NE} is required. Instead, realistic market behavior of the participants is sufficient for evaluation.
\par 
We performed all our training runs on a DGX-1 with 80 cores. Still, we reached its performance limits several times, mainly because of the \ac{MADDPG} experiments with hundreds of thousands of \ac{OPF} calculations. This further reinforces the importance of this kind of research to use domain knowledge to speed up agent training. 




\bibliographystyle{elsarticle-num.bst} 
\bibliography{bibliography.bib}

\end{document}